\RequirePackage{etex}
\documentclass[a4paper,11pt]{article}
\usepackage{jcappub}
\pdfoutput=1
\usepackage{paralist}
\usepackage{graphicx}
\usepackage{amsfonts}
\usepackage{amssymb}
\usepackage{xcolor}
\usepackage[export]{adjustbox}
\usepackage{amsmath}
\usepackage{slashed}
\usepackage{dcolumn}
\usepackage{verbatim}
\usepackage{float}
\usepackage{multirow}
\usepackage{soul}
\usepackage{xspace}
\usepackage[normalem]{ulem}
\usepackage{hyperref}
\usepackage{tabularx}
\usepackage{setspace}

\newcommand{\beq}{\begin{equation}}
\newcommand{\eeq}{\end{equation}}
\newcommand{\bea}{\begin{eqnarray}}
\newcommand{\eea}{\end{eqnarray}}

\usepackage{pdfbase}[2017/03/16]
\usepackage{xparse,ocgbase}
\usepackage{xcolor,calc}
\usepackage{tikzpagenodes,linegoal}
\usetikzlibrary{calc}
\usepackage{tcolorbox}

\ExplSyntaxOn
\let\tpPdfLink\pbs_pdflink:nn
\let\tpPdfAnnot\pbs_pdfannot:nnnn\let\tpPdfLastAnn\pbs_pdflastann:
\let\tpAppendToFields\pbs_appendtofields:n
\def\tpPdfXform{\pbs_pdfxform:nnnnn{1}{1}{}{}}
\let\tpPdfLastXform\pbs_pdflastxform:
\let\cListSet\clist_set:Nn\let\cListItem\clist_item:Nn
\ExplSyntaxOff

\usepackage{pdfbase}[2017/03/16]
\usepackage{xparse,ocgbase}
\usepackage{xcolor,calc}
\usepackage{tikzpagenodes,linegoal}
\usetikzlibrary{calc}
\usepackage{tcolorbox}

\ExplSyntaxOn
\let\tpPdfLink\pbs_pdflink:nn
\let\tpPdfAnnot\pbs_pdfannot:nnnn\let\tpPdfLastAnn\pbs_pdflastann:
\let\tpAppendToFields\pbs_appendtofields:n
\def\tpPdfXform{\pbs_pdfxform:nnnnn{1}{1}{}{}}
\let\tpPdfLastXform\pbs_pdflastxform:
\let\cListSet\clist_set:Nn\let\cListItem\clist_item:Nn
\ExplSyntaxOff

\makeatletter
\NewDocumentCommand{\tooltip}{%
  ssssO{\ifdefined\@linkcolor\@linkcolor\else blue\fi}mO{yellow!20}mO{0pt,0pt}%
}{{%
  \leavevmode%
\IfBooleanT{#2}{%
    for variants with two and more stars, put tip box on a PDF Layer (OCG)
    \ocgbase@new@ocg{tipOCG.\thetcnt}{%
      /Print<</PrintState/OFF>>/Export<</ExportState/OFF>>%
    }{false}%
    \xdef\tpTipOcg{\ocgbase@last@ocg}%
    \ocgbase@add@ocg@to@radiobtn@grp{tool@tips}{\ocgbase@last@ocg}%
  }%
  \tpPdfLink{%
    \IfBooleanTF{#4}{%
      /Subtype/Link/Border[0 0 0]/A <</S/SetOCGState/State [/Toggle \tpTipOcg]>>
    }{%
      /Subtype/Screen%
      /AA<<%
        \IfBooleanTF{#3}{%
          /E<</S/SetOCGState/State [/Toggle \tpTipOcg]>>%
        }{%
          \IfBooleanTF{#2}{%
            /E<</S/SetOCGState/State [/ON \tpTipOcg]>>%
            /X<</S/SetOCGState/State [/OFF \tpTipOcg]>>%
          }{
            \IfBooleanTF{#1}{%
              /E<</S/JavaScript/JS(%
                var fd=this.getField('tip.\thetcnt');%
                if(typeof(click\thetcnt)=='undefined'){%
                  var click\thetcnt=false;%
                  var fdor\thetcnt=fd.rect;var dragging\thetcnt=false;%
                }%
                if(fd.display==display.hidden){%
                  fd.delay=true;fd.display=display.visible;fd.delay=false;%
                }else{%
                  if(!click\thetcnt&&!dragging\thetcnt){fd.display=display.hidden;}%
                  if(!dragging\thetcnt){click\thetcnt=false;}%
                }%
                this.dirty=false;%
              )>>%
            }{%
              /E<</S/JavaScript/JS(%
                var fd=this.getField('tip.\thetcnt');%
                if(typeof(click\thetcnt)=='undefined'){%
                  var click\thetcnt=false;%
                  var fdor\thetcnt=fd.rect;var dragging\thetcnt=false;%
                }%
                if(fd.display==display.hidden){%
                  fd.delay=true;fd.display=display.visible;fd.delay=false;%
                }%
               this.dirty=false;%
              )>>%
              /X<</S/JavaScript/JS(%
                if(!click\thetcnt&&!dragging\thetcnt){fd.display=display.hidden;}%
                if(!dragging\thetcnt){click\thetcnt=false;}%
                this.dirty=false;%
              )>>%
            }%
            /U<</S/JavaScript/JS(click\thetcnt=true;this.dirty=false;)>>%
            /PC<</S/JavaScript/JS (%
              var fd=this.getField('tip.\thetcnt');%
              try{fd.rect=fdor\thetcnt;}catch(e){}%
              fd.display=display.hidden;this.dirty=false;%
            )>>%
            /PO<</S/JavaScript/JS(this.dirty=false;)>>%
          }%
        }%
      >>%
    }%
  }{{\color{#5}#6}}%
  \sbox\tiptext{%
    \IfBooleanT{#2}{%
      \ocgbase@oc@bdc{\tpTipOcg}\ocgbase@open@stack@push{\tpTipOcg}}%
    \tcbox[colframe=black,colback=#7,size=fbox,arc=1ex,sharp corners=southwest]{#8}%
    \IfBooleanT{#2}{\ocgbase@oc@emc\ocgbase@open@stack@pop\tpNull}%
  }%
  \cListSet\tpOffsets{#9}%
  \edef\twd{\the\wd\tiptext}%
  \edef\tht{\the\ht\tiptext}%
  \edef\tdp{\the\dp\tiptext}%
  \tipshift=0pt%
  \IfBooleanTF{#2}{%
    \setlength\whatsleft{\linegoal}%
  }{%
    \measureremainder{\whatsleft}%
  }%
  \ifdim\whatsleft<\dimexpr\twd+\cListItem\tpOffsets{1}\relax%
    \setlength\tipshift{\whatsleft-\twd-\cListItem\tpOffsets{1}}\fi%
  \IfBooleanF{#2}{\tpPdfXform{\tiptext}}%
  \raisebox{\heightof{#6}+\tdp+\cListItem\tpOffsets{2}}[0pt][0pt]{%
    \makebox[0pt][l]{\hspace{\dimexpr\tipshift+\cListItem\tpOffsets{1}\relax}%
    \IfBooleanTF{#2}{\usebox{\tiptext}}{%
      \tpPdfAnnot{\twd}{\tht}{\tdp}{%
        /Subtype/Widget/FT/Btn/T (tip.\thetcnt)%
        /AP<</N \tpPdfLastXform>>%
        /MK<</TP 1/I \tpPdfLastXform/IF<</S/A/FB true/A [0.0 0.0]>>>>%
        /Ff 65536/F 3%
        /AA <<%
          /U <<%
            /S/JavaScript/JS(%
              var fd=event.target;%
              var mX=this.mouseX;var mY=this.mouseY;%
              var drag=function(){%
                var nX=this.mouseX;var nY=this.mouseY;%
                var dX=nX-mX;var dY=nY-mY;%
                var fdr=fd.rect;%
                fdr[0]+=dX;fdr[1]+=dY;fdr[2]+=dX;fdr[3]+=dY;%
                fd.rect=fdr;mX=nX;mY=nY;%
              };%
              if(!dragging\thetcnt){%
                dragging\thetcnt=true;Int=app.setInterval("drag()",1);%
              }%
              else{app.clearInterval(Int);dragging\thetcnt=false;}%
              this.dirty=false;%
            )%
          >>%
        >>%
      }%
      \tpAppendToFields{\tpPdfLastAnn}%
    }%
  }}%
  \stepcounter{tcnt}%
}}
\makeatother
\newsavebox\tiptext\newcounter{tcnt}
\newlength{\whatsleft}\newlength{\tipshift}
\newcommand{\measureremainder}[1]{%
  \begin{tikzpicture}[overlay,remember picture]
    \path let \p0 = (0,0), \p1 = (current page.east) in
      [/utils/exec={\pgfmathsetlength#1{\x1-\x0}\global#1=#1}];
  \end{tikzpicture}%
}

\newcommand{\msun}{{\rm M}_\odot}

\renewcommand{\paragraph}[1]{~\\ \noindent{\bf \emph{#1} --}}

\newcommand{\vtwo}[1]{\textcolor{black}{#1}}
\newcommand{\vthree}[1]{\textcolor{black}{#1}}

\DeclareRobustCommand{\okina}{%
  \raisebox{\dimexpr\fontcharht\font`A-\height}{%
    \scalebox{0.8}{`}%
  }%
}
\allowdisplaybreaks

\begin{document}

\title{Dark Dwarfs:~\\ Dark Matter-Powered Sub-Stellar Objects Awaiting Discovery at the Galactic Center}

\author[a,1]{Djuna Croon, \note{ORCID:~\href{https://orcid.org/0000-0003-3359-3706}{0000-0003-3359-3706}}} 
 \emailAdd{djuna.l.croon@durham.ac.uk}
 
\affiliation[a]{Institute for Particle Physics Phenomenology, Department of Physics, Durham University, Durham DH1 3LE, U.K.}

\author[b,2]{Jeremy Sakstein, \note{ORCID:~\href{https://orcid.org/0000-0002-9780-0922}{0000-0002-9780-0922}}}
\affiliation[b]{Department of Physics \& Astronomy, University of Hawai\okina i, Watanabe Hall, 2505 Correa Road, Honolulu, HI, 96822, USA}
\emailAdd{sakstein@hawaii.edu}

\author[c,3]{Juri Smirnov, \note{ORCID:~\href{http://orcid.org/0000-0002-3082-0929}{0000-0002-3082-0929}}}
\emailAdd{{juri.smirnov@liverpool.ac.uk}}

\affiliation[c]{Department of Mathematical Sciences, University of Liverpool,
Liverpool, L69 7ZL, United Kingdom}
\author[a]{and Jack Streeter}

\date{\today}

\abstract{
We investigate the effects of dark matter annihilation on objects with masses close to the sub-stellar limit, finding that the minimum mass for stable hydrogen burning is larger than the $\sim0.075\msun$ value predicted in the Standard Model.~Below this limit, cooling brown dwarfs evolve into stable dark matter-powered objects that we name \textit{dark dwarfs}.~The timescale of this transition depends on the ambient dark matter density $\rho_{\rm DM}$ and circular velocity $v_{\rm DM}$ but is independent of the dark matter mass.~We predict a population of dark dwarfs close to the galactic center, where the dark matter density is expected to be $\rho_{\rm DM}\gtrsim 10^{3}$ GeV/cm$^3$.~At larger galactic radii the dark matter density is too low for these objects to have yet formed within the age of the universe.~Dark dwarfs  retain their initial lithium-7 in mass ranges where brown/red dwarfs would destroy it, providing a method for detecting them.
}

\subheader{IPPP/24/52,     LTH1382}

\maketitle

\section{Introduction}

Discovering the microphysical description of dark matter (DM) is a paramount goal of particle physics and cosmology.~All of our current evidence for DM's existence is indirect, and we have no information pertaining to its mass and interactions with visible matter or other dark sector particles.~We do however know that  DM  needs to be produced in the early universe.~Among the many mechanisms for this, thermal freeze-out is a well-motivated minimal scenario.~Provided that Standard Model (SM) particles are present in the final states of the process that sets the relic density, there is an energy transfer to the visible sector~\cite{Steigman:1984ac, Jungman:1995df, Bertone:2004pz, Bertone:2016nfn}.~This process can be detected via annihilation signals in space~\cite{Geringer-Sameth:2014yza,Fermi-LAT:2015att}, or scattering in the laboratory~\cite{Smirnov:2020zwf, Parikh:2023qtk}, but also affects astrophysical objects~\cite{Batell:2009zp, Pospelov:2007mp, Rothstein:2009pm, Pospelov:2008jd,Chen:2009ab,Schuster:2009au,Schuster:2009fc,Bell_2011,Feng:2015hja,Kouvaris:2010,Feng:2016ijc,Allahverdi:2016fvl,Leane:2017vag,Arina:2017sng,Albert:2018jwh,Albert:2018vcq,Nisa:2019mpb,Niblaeus:2019gjk,Cuoco:2019mlb,Serini:2020yhb,Acevedo:2020gro,Mazziotta:2020foa,Bell:2021pyy,Bose:2021cou,Smirnov:2022zip,Croon:2023trk,John:2024thz}.

Objects that are not supported by nuclear burning such as white dwarfs, neutron stars~\cite{Goldman:1989nd,
Gould:1989gw,
Kouvaris:2007ay,
Bertone:2007ae,
deLavallaz:2010wp,
Kouvaris:2010vv,
McDermott:2011jp,
Kouvaris:2011fi,
Guver:2012ba,
Bramante:2013hn,
Bell:2013xk,
Bramante:2013nma,
Bertoni:2013bsa,
Kouvaris:2010jy,
McCullough:2010ai,
Perez-Garcia:2014dra,
Bramante:2015cua,
Graham:2015apa,
Cermeno:2016olb,
Graham:2018efk,
Acevedo:2019gre,
Janish:2019nkk,
Krall:2017xij,
McKeen:2018xwc,
Baryakhtar:2017dbj,
Raj:2017wrv,
Bell:2018pkk,
Chen:2018ohx,
Garani:2018kkd,
Dasgupta:2019juq,
Hamaguchi:2019oev,
Camargo:2019wou,
Bell:2019pyc,
Acevedo:2019agu,
Joglekar:2019vzy,
Joglekar:2020liw,
Bell:2020jou,
Garani:2020wge,
Leane:2021ihh,Bose:2021yhz,Collier:2022cpr}, planets~\cite{Mack:2007xj, Bramante:2019fhi, Leane:2021tjj,Croon:2023bmu}, and brown dwarfs~\cite{Leane:2020wob,Leane:2021ihh} have been shown to exhibit observable signatures based on DM interactions, as an additional energy injection can lead to changes in their structure and evolution. \vtwo{Note that large fractions of the dark matter parameter space can lead to substantial capture rates in celestial bodies.~For example, in the case of spin-dependent proton scattering, cross sections above $\sim 10^{-25} \, \rm{cm}^2$ are practically unconstrained~\cite{QUEST-DMC:2025qsa}, with some exceptions which depend on modeling assumptions e.g., cosmology dependent constraints from the CMB \cite{Boddy:2018kfv}, the lack of MW satellites \cite{Nadler:2019zrb} for velocity-independent scattering cross sections, and the existence of Jupiter \cite{Croon:2023bmu}.}~\vthree{In particular, the mass range in the GeV region at large cross sections is not constrained by Earth and Mars heating~\cite{Bramante:2019fhi} because evaporation would eject dark matter particles in that mass range from rocky planets but would retain them in gas giants and brown dwarfs~\cite{Garani:2021feo}.~Such cross-sections} would lead to geometric capture rates in brown dwarfs, containing predominantly hydrogen nuclei.~Thus, masses above the evaporation threshold at the GeV scale are testable in heating scenarios.~Note that these objects are also excellent probes of modified theories of gravity \cite{Sakstein:2015zoa,Sakstein:2015aac}. 

\begin{figure}[t!]
    \centering
    \hspace*{-.5cm}
\includegraphics[width=.85\linewidth]{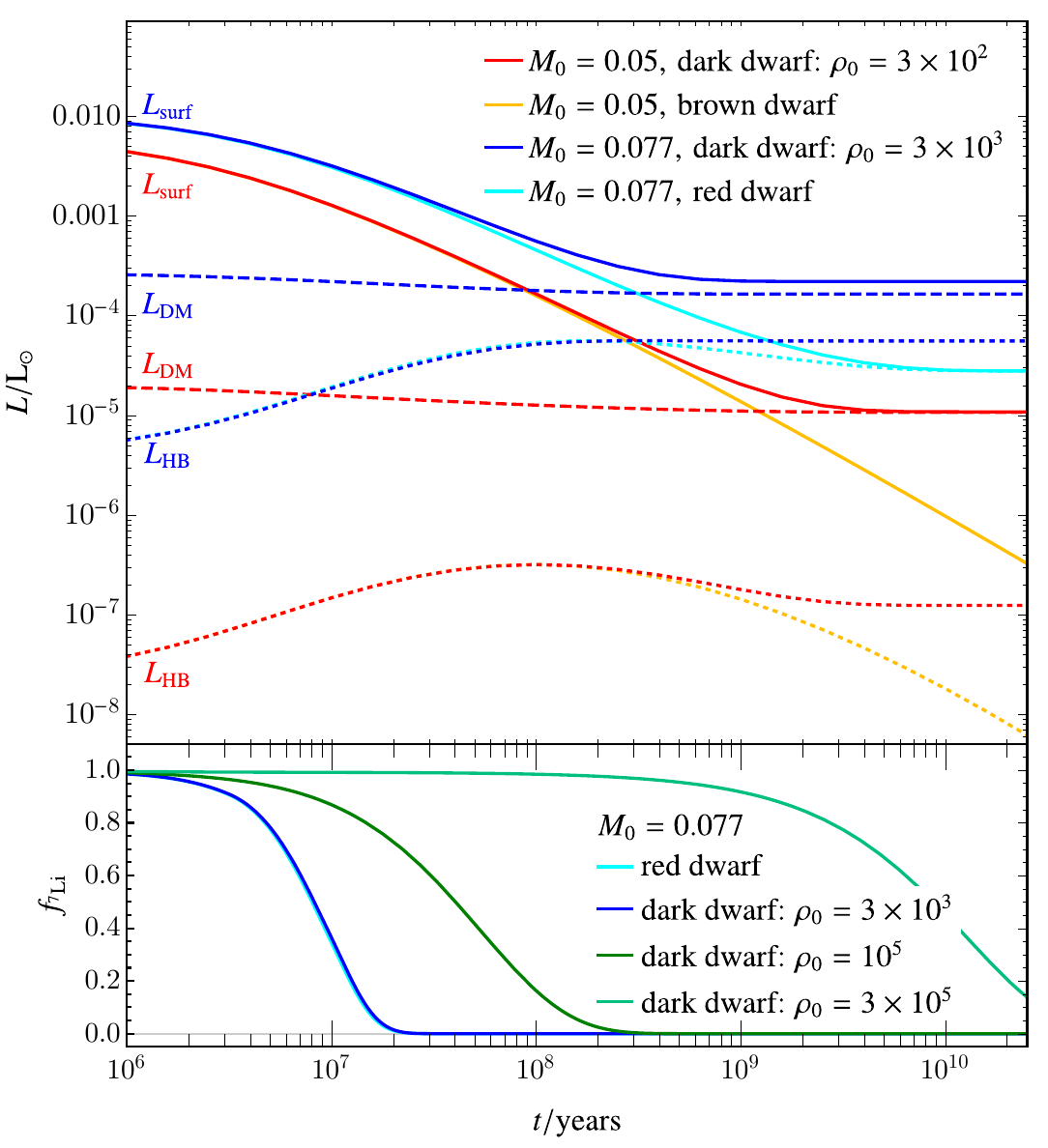}
    \caption{Effects of DM energy injection on luminosity and lithium depletion for objects with masses $M_0\equiv M/M_\odot $ \vtwo{embedded in DM density $\rho_0\equiv \rho_{\rm DM} / (\rm GeV \, cm^{-3})$}.~\textbf{Upper panel:}~time-evolution of the total luminosity $L_{\rm surf}$ (solid lines), luminosity from DM burning $L_{\rm DM}$ (dashed lines) and luminosity in hydrogen burning (dotted lines).~\textbf{Lower panel:}~time evolution of the surviving fraction of $^7$Li.~Dark dwarfs retain their initial lithium-7 while red dwarfs deplete it.}
    \label{fig:intro}
\end{figure}

In this work we investigate the effects of energy injection from DM annihilation on celestial bodies at the stellar mass boundary.~These objects, which have masses $M\lesssim0.1\msun$, are too light to fuse hydrogen on the PPI chain due to their cool core temperatures $T\sim 10^6$~K.~Those heavier than $\sim$$0.075\msun$, red dwarfs or M-dwarfs, evolve into equilibrium states that fuse hydrogen to helium-3.~ Lighter bodies, so called brown dwarfs, are unable to do so.~They experience transitory periods of hydrogen, lithium, and deuterium burning, but ultimately cool and contract eternally.

As shown in Fig.~\ref{fig:intro} the situation is dramatically different when DM annihilation is present.~Contracting objects attain an equilibrium configuration where they are supported primarily by DM heating.~This steady-state exists for any non-zero DM density, but is attained in a Hubble time for $\rho_{\rm DM} \gtrsim 10^3 \rm GeV \, cm^{-3}$ \vthree{(for reference, assuming an NFW profile, the DM density a parsec away from the Galactic Centre (GC) is $\rho_{\rm DM} =10^4 \, \rm GeV \, cm^{-3} $)}.~We name these objects \textit{Dark Dwarfs} (DDs), and differentiate them from red dwarfs near the hydrogen burning limit --- $M\sim0.075\msun$ in the SM --- which are supported by a combination of nuclear burning and DM heating.~DDs are physically distinct from brown/red dwarfs in several ways:
\begin{compactitem}
    \item They are predominantly powered by DM heating, exhibiting a component of stable hydrogen burning.
    \item Their luminosity, radius, and effective temperature are constant in time.
    \item Lithium depletion occurs at larger threshold masses than the SM prediction.
\end{compactitem}
These properties, which are exemplified in Fig.~\ref{fig:intro}, are independent of the DM mass but do depend upon the ambient DM density and velocity dispersion.\vtwo{~Moreover, through our novel analytic model we demonstrate that DM heating leads to non-linear feedback that affects the energy emission for objects near the $0.075\msun$ boundary, as can be seen in Fig.~\ref{fig:intro}.~A consequence of this is that}~DDs can be identified by their enhanced lithium abundance despite a relatively large mass, and old stellar age.

\section{Analytic Model for Dark Dwarfs}
\label{sec:analytic}
Sub-stellar objects are relatively simple compared with more massive bodies.~Their homogeneous structure and lack of strong nuclear burning make them well-described by analytic models.~They are supported by a combination of the degeneracy pressure of the electrons and the gas pressure of the ions.~Under these assumptions the equation of state (EOS) is that of an $n=3/2$ polytrope except near the surface where there is a transition to a phase of molecular hydrogen \cite{2016AdAst2016E..13A}.~The full details of this model are reviewed in the Supplementary Material.

For our purposes, it is sufficient to begin with the model's prediction for the surface luminosity:
\begin{equation}
    \frac{L_{\rm surf}}{L_\odot}= \frac{0.0578353 b_1^3 M_0}{\kappa_R} \psi ^{3 \nu } \left(\frac{M_0^{5/3} \psi ^{-\nu }}{b_1 \kappa_R (1+\gamma+\alpha \psi  )^2}\right)^{1/7}.
\end{equation}
where $M_0 = M/M_\odot$, $\kappa_R$ is the Rosseland mean opacity at the photosphere (we set $\kappa_R=0.01$ cm$^2$/g in what follows, typical for these objects), $b_1$ and $\nu$ are parameters of the molecular hydrogen EOS
(we shall use parameter point D in the numerical calculations in this work, defined in table~\ref{tab:models} in the Supplemental Material, with $b_1=2$ and $\nu =1.6$), $\psi = k_B T/E_F$ is the degeneracy parameter, and $\gamma$ is a function of $\psi$ given in Eq.~\eqref{eq:gammaeq}.~The properties of the star are completely determined once a value of $\psi$ is specified.~The steady-state condition that determines $\psi$ is energy conservation --- the surface luminosity must balance the luminosity from all heat sources:
\begin{equation}
\label{eq:energycons}
    L_{\rm surf} = \sum_i L_{i}
\end{equation}
where $i$ runs over all processes injecting energy into the star, which in our case include nuclear burning and dark matter annihilation.~The polytropic approximation ceases to be valid in very low mass objects because Coulomb scattering begins to become important \cite{1983psen.book.....C,Saumon:1995bn,burrows1993science}.~This breakdown happens for \vtwo{$M\lesssim {0.018}\msun$} in the SM and, as derived in Appendix~\ref{app:polytropic_approximation_failure}, breaks down at lower masses when the DM energy injection is important.~The $n=3/2$ approximation remains valid for all objects studied in this work.

\begin{figure*}[t]
    \centering
    \includegraphics[width=0.45\textwidth]{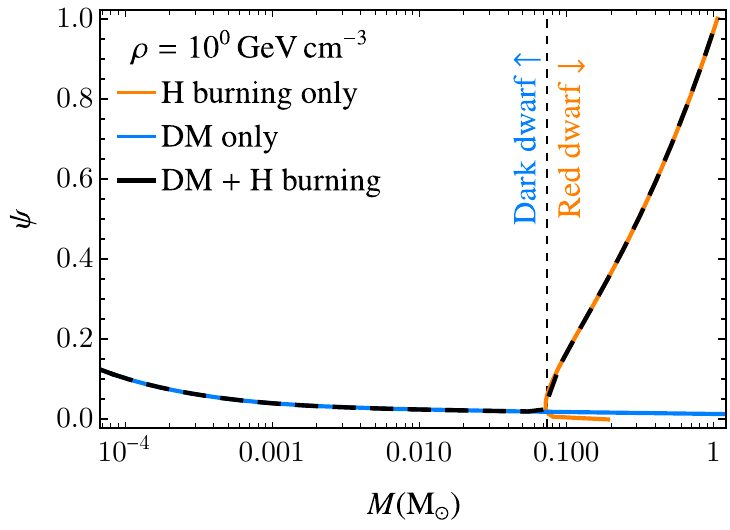}
    \includegraphics[width=0.45\textwidth]{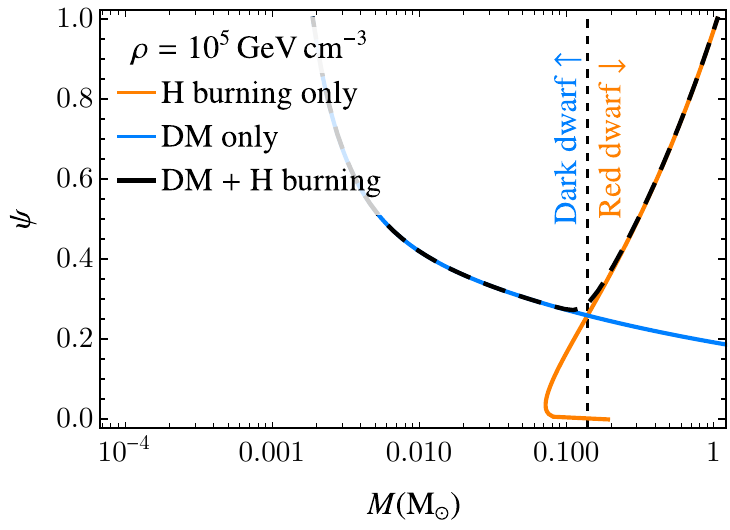}
    \caption{Degeneracy parameter $\psi$ as a function of mass for stars supported by both hydrogen burning and DM annihilation.~The blue curves show solutions where $L_{\rm surf}=L_{\rm DM}$ (including gravitational focusing), corresponding to stars supported solely by DM annihilation;~the orange curve shows solutions for the SM case where objects are supported solely by hydrogen burning, $L_{\rm surf}=L_{\rm HB}$;~and the black dashed lines show the solution for the general case $L_{\rm surf}=L_{\rm DM}+L_{\rm HB}$.~Lighter objects are primarily DM supported (dark dwarfs) whereas heavier masses behave like SM H-burning stars (red dwarfs).}
    \label{fig:M_psi_hybrid}
\end{figure*}

Before investigating the effects of DM, it is instructive to review SM objects, which are supported solely by hydrogen burning.~The central temperatures are not sufficient to fuse $^3$He to $^4$He so the PP-chain is only comprised of the reactions
\begin{align}
    \textrm{p}+\textrm{p}&\rightarrow \textrm{d} + \textrm{e}^+ + \nu_e\\
    \textrm{p} + \textrm{d} &\rightarrow ^3\!\!{\rm He}+\gamma.
\end{align}
The luminosity from these processes, assuming the $n=3/2$ polytrope, is \cite{2016AdAst2016E..13A}
\begin{equation}
    \frac{L_{\rm HB}}{L_\odot}=7.33\times10^{16}M_0^{11.977}\frac{\psi^{6.0316}}{(1+\gamma(\psi)+\alpha\psi)^{16.466}},
    \label{eq:LHB}
\end{equation}
which was calculated by integrating the energy generation rates over the volume of the star.~At the core temperatures and densities relevant for the objects we study, the majority of the thermonuclear energy is produced from deuterium burning.~Imposing energy conservation, stable hydrogen burning is achieved whenever $L_{\rm HB}=L_{\rm surf}$, which yields 
\begin{equation}
 M_0 = 0.03362 \, b_1^{0.26605}
\frac{(1+\gamma(\psi)+\alpha\psi)^{1.5069}}{\psi^{0.266053 \nu -0.56165}}\,.
    \label{eq:M1HB}
\end{equation}
There is a minimum value of $\psi$ for which Eq.~\eqref{eq:M1HB} has solutions.~Below this, the object cannot burn hydrogen stably because the core temperatures and densities are too low to sustain a sufficient nuclear burning rate to balance the surface luminosity.~This boundary corresponds to the minimum mass for hydrogen burning (MMHB).~Stars heavier than this are red dwarfs while lighter objects are brown dwarfs.~In our model, the specific value is $M_{\rm MMHB}=0.075\msun$, consistent with the literature e.g., \cite{burrows1993science,2016AdAst2016E..13A}.

To calculate the effects of DM energy injection, we will work in the limit of capture-annihilation equilibrium where the amount of DM captured by the object precisely balances that lost to annihilation.~This equilibrium is reached on time scales of $\sim$Myr and shorter assuming DM annihilation rates compatible with thermal freezeout (see the appendix of Ref.~\cite{Leane:2020wob} for a detailed discussion of this).~We assume that the energy from the DM injection reaches the surface without delay.~This is justified because the objects we consider are fully convective with convection time-scales of order years~\cite{Freytag:2010kn} and velocities $v_{\rm conv} \sim 10^4 $ cm/s (see Fig.~10 of~\cite{Freytag:2010kn}), the time to reach the surface is then $\sim$ 100 days, far shorter than the cooling timescale of the object.~\vtwo{The energy transport properties of the star are not affected by the presence of DM annihilation because this process drives radiative regions to shrink in favor of larger convection zones \cite{2011MNRAS.410..535C,John:2024thz,Scott:2008ns} and, as noted above, these stars are already fully-convective.}~\vthree{In addition, DM itself can act as a source of heat transport within the star \cite{Gould:1989hm,Scott:2008ns,2009ApJ...705..135C,Lopes:2001ig,Lopes:2002gp,Taoso:2010tg}, but its contribution is negligible due to the small DM abundance relative to the SM matter.}

\vthree{Under the assumptions above}, the DM heat injection rate is given at any point of the stellar evolution by \cite{Leane:2020wob, Leane:2023woh, Croon:2023trk}
\begin{equation}
\label{eq:LDM}
\begin{split}
    L_{\rm DM}&=m_{\rm DM}\pi R^2\Phi,  \quad \text{with}
    \\
    \Phi &=\sqrt{\frac{8}{3\pi}}\frac{\rho_{\rm DM}v_{\rm DM}f_{\rm cap}}{m_{\rm DM}} \left( 1+ \frac{3}{2} \frac{v_{\rm esc}^2}{v_{\rm DM}^2}\right)
\end{split}
\end{equation}
where $v_{\rm esc}^2=2GM/R$ is the escape velocity, $\rho_{\rm DM}$ is the ambient DM density, $v_{\rm DM}$ is circular velocity of galactic DM, $f_{\rm cap}$ is the fraction of DM that is captured, and $m_{\rm DM}$ is the DM mass;~the term in the brackets accounts for gravitational focusing.~Throughout this work, we fix $v_{\rm DM}=50$ km/s as a conservative benchmark \vtwo{for the GC~\cite{Karukes:2019jwa,Lin:2019yux}}.~These equations have an important consequence:~the DM luminosity is independent of the DM mass.~Since the properties of the object are determined solely by energy conservation, the characteristics of objects supported partially or fully by DM annihilation are similarly DM mass-independent.~\vthree{Similarly, our results are insensitive to the details of the DM profile because the total luminosity, found by integrating over this, is determined by the star's radius and the DM flux $\Phi$.~This implies that scenarios where there is a substantial amount of DM at the star's surface \cite{Leane:2022hkk} are also covered by our analysis. }

\vthree{We will set the fraction of dark matter captured, $f_{\rm cap}$, to unity in what follows.~This translates into an assumption about mass and cross section which can be read of from Fig. S5 in \cite{Leane:2020wob}.~However, we note that the observational signals which we predict can be observable when only a subfraction of dark matter is captured i.e., $f_{\rm cap} <1$ because the dark matter density near the galactic center exceeds the local value by several orders of magnitude.~In our plots below, we show results for varying $\rho_{\rm DM}$ assuming $f_{\rm cap} =1$, however, they simply rescale as $\rho_{\rm DM} f_{\rm cap}$ for scenarios where dark matter capture is not 100\% efficient.}~Another scenario that is accessible in high dark matter density environments is a setup where a subfraction of dark matter exhibits large elastic cross sections \vthree{as expected in models with composite dark matter candidates~\cite{DeLuca:2018mzn,Gross:2018zha}. Here, our results apply by replacing the capture fraction $f_{\rm cap}$ of dark matter in the object by the total fraction of the strongly interacting dark matter sub-component in the halo $f_{\rm strong}$. This fraction can be sizable depending on the dark matter mass, and is largely untested by terrestrial searches, see for example Ref.~\cite{McKeen:2022poo}.}

To understand the differences between hydrogen and DM burning, it is enlightening to consider a star supported entirely by DM annihilation, i.e.,~$L_{\rm DM}=L_{\rm surf}$.~Using equation~\eqref{eq:Rpoly} for the radius in \eqref{eq:LDM}, we can find an analytic solution when gravitational focusing is neglected:
\begin{equation}
\begin{split}
    M_0 =&8.58\times10^{-7}
    \times \frac{(1+\gamma(\psi)+\alpha\psi)^{1.19895}}{\psi^{1.4993 \nu}}
    \\ &\times 
\left(
    \frac{\rho_{\rm DM}}{\rm GeV \, cm^{-3}}
    \frac{v_{\rm DM}}{50\rm km \, s^{-1}}
    \frac{f_{\rm cap}}{b_1^{2.857}}\right)^{0.52475}
.
\end{split}
\label{eq:M1dm}
\end{equation}
In contrast to the equivalent formula for hydrogen-supported objects \eqref{eq:M1HB}, Eq.~\eqref{eq:M1dm} has no minimum --- DM can support structures of arbitrarily small masses.~This is because, unlike hydrogen burning, the DM burning rate does not depend on the stellar properties --- DM burning is ever-present.~Including gravitational focusing does not alter this conclusion and in fact produces stronger effects.~It is included in our numerical studies below.

Fig.~\ref{fig:M_psi_hybrid} shows numerical solutions for general cases where stars are heated by both, hydrogen and dark matter annihilation, i.e.~$L_{\rm surf}=L_{\rm DM}+L_{\rm HB}$.~Evidently, the low mass solutions are \emph{dark dwarfs}, supported almost entirely by DM annihilation, whereas the high mass solutions are similar to  red dwarf stars --- they are supported almost entirely by hydrogen burning but have some amount of DM support.~In the transition region, both sources of burning are important.~We therefore expect that, unlike in the SM, all brown dwarfs will ultimately evolve into dark dwarfs.~Whether this happens within the age of the universe depends on the incident DM flux and circular velocity.

\section{Time Evolution and Stability}
To study the time evolution that leads to the formation of dark dwarfs, we use the cooling model introduced in \cite{stevenson1991search,burrows1993science}.~Using the first law of thermodynamics, the equation for the energy of a contracting
star is 
\begin{equation}
    T\frac{\mathrm{d} S}{\mathrm{d} t}=\varepsilon-\frac{\partial L_{\rm surf}}{\partial M}
\end{equation}
where $S$ is the entropy per unit mass and $\varepsilon$ is the energy generation rate per unit mass from burning.~Integrating this equation, employing the equation of state for molecular hydrogen, and using the EOS and properties of $n=3/2$ polytropes one finds \cite{2016AdAst2016E..13A}
\begin{equation}
\label{eq:psi_deriv}
    \frac{\mathrm{d}\psi}{\mathrm{d} t} = \frac{\bar{\mu}}{1.5\mu_e^{\frac83}E_0}M_0^{{-7/3}}
    (1+\gamma(\psi)+\alpha\psi)^2(L_{\rm burn}-L_{\rm surf}),
\end{equation}
where $E_0=6.73857\times10^{49}$ erg, $L_{\rm burn}$ is the energy in burning, and
\begin{equation}
\label{eq:bar_mu}
    \frac{1}{\bar{\mu}}=\frac{1}{{\mu}}+\frac{3X_{H^+}(1-X_{H^+})}{2(2-X_{H^+})}
\end{equation}
with $X_{H^+}$ the mass fraction of ionized hydrogen;~$\bar{\mu}=1.02$ for parameter point D, see the Supplementary Material.~In the absence of any burning, equation \eqref{eq:psi_deriv} implies that $\psi$ will decrease from its initial value i.e., that over time $k_B T$ will decrease relative to \vtwo{the Fermi energy} $E_F$ (see equation~\eqref{eq:psi_def}) and the star will be increasingly supported by degeneracy pressure.~In the presence of burning, there is a steady-state when the right hand side of equation \eqref{eq:psi_deriv} is zero i.e., when $L_{\rm burn}=L_{\rm surf}$.~These are the solutions we derived above.~We thus expect that any initial configuration will evolve towards this steady-state, reaching it at sufficiently late times.~

The stability of steady-state solutions can be determined as follows.~Letting $\psi_0$ be the value of $\psi$ where the steady-state is achieved, we can write $\psi(t)=\psi_0+\delta\psi(t)$ and Taylor-expand equation \eqref{eq:psi_deriv} to find $\dot{\delta\psi}=f(\psi_0)\delta\psi+\mathcal{O}(\delta\psi^2)$, where $f(\psi_0)$ is the derivative of the right hand side of \eqref{eq:psi_deriv} with respect to $\psi$ and evaluated at $\psi_0$.~If $f(\psi_0)<0$ then the equilibrium is stable.~The resulting expressions are long and not informative so we do not give them here.~They can be found in our accompanying  
~\href{https://zenodo.org/records/13141908}{code}.~Using the same code, we found that dark dwarfs are always stable.~This is true for objects supported by a combination of H-burning and DM burning, or objects supported solely by DM burning.~In contrast, only the branch of SM stars (solely H-burning) with $\psi>\psi_{\rm min}$ is stable.~

\begin{figure}
    \centering
    \vspace{-5mm}    \includegraphics[width=0.85\linewidth]{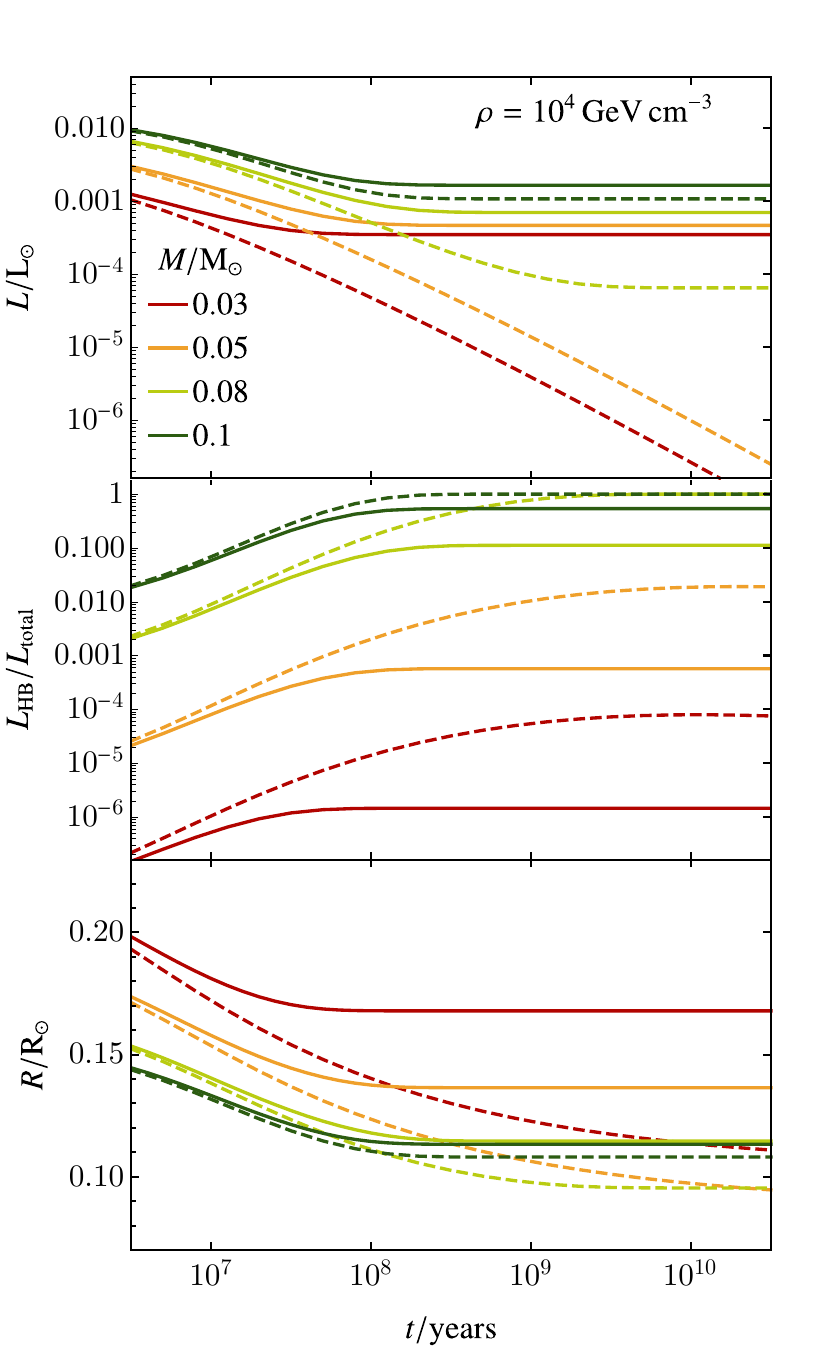}
    \caption{
  Time evolution of important stellar properties.~The continuous lines correspond to objects evolving in the presence of DM annihilation while dashed lines indicate the SM predictions.~Four representative stellar masses indicated in the top panel are shown.\vtwo{~In the SM, the lighter two objects evolve to become brown dwarfs, and the heavier objects evolve to become red dwarfs, as can be observed by the continuous decrease versus stabilizing of the luminosity at late times.}
  }

    \label{fig:cooling_L_vs_t}
\end{figure}

\begin{figure}[b]
    \centering
    \includegraphics[width=0.85\linewidth]{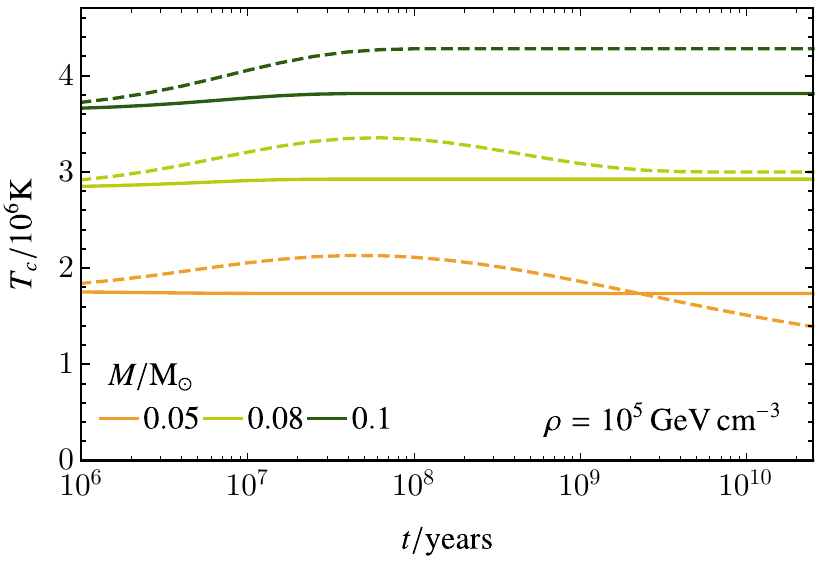}
    \caption{Time evolution of the core temperatures of \vtwo{three} objects with different masses with (solid lines) and without (dashed lines) DM energy injection.~Coulomb corrections to the pressure are included.\vtwo{~In the SM, the lighter objects evolve to become a brown dwarf, and the heavier objects evolve to become red dwarfs.}
    }
    \label{fig:Tcplot5}
\end{figure}

In Fig.~\ref{fig:cooling_L_vs_t} we show  properties of objects in the presence of DM annihilation as a function of time found by solving Eq.~\eqref{eq:psi_deriv}.~\vtwo{At early times, the evolution of the objects are dominated by gravitational collapse so have similar properties.~At later times, s}everal differences between SM objects and those with some amount of DM burning are evident.~SM brown dwarfs lighter than the MMHB cool continuously, contracting and becoming dimmer.~SM objects heavier than the MMHB cool and evolve to red dwarfs where they are supported by H-burning.~In contrast,  in the presence of DM burning lighter objects cool similarly to brown dwarfs until they reach the steady-state dark dwarf solution with constant core temperature, radius, and brightness.~The time to reach the dark dwarf state is shorter for larger ambient DM densities.~The MMHB is also DM density dependent.

We found that for typical galactic center DM velocities and $\rho\gtrsim 10^3  \rm \, GeV \, cm^{-3}$ objects with mass $ M \lesssim 0.05 \msun$ begin to  display differences from brown dwarfs (dashed lines) on a timescale shorter than a Hubble time.~Objects in lower density environments will not reach the steady-state within the age of the universe and will appear as brown dwarfs for all intents and purposes.~Heavier objects will evolve into red dwarfs but given the same mass their properties are different from the SM prediction due to a non-negligible contribution from DM annihilation.~At higher DM densities $\sim 10^{5} \rm \, GeV \, cm^{-3}$, even for higher mass objects  there is a significant gap between the luminosity output of the star and the fraction supplied by hydrogen burning.~As we will discuss in the next section, light element burning is also affected leading to a potential search strategy via specific spectroscopic markers.

\section{Lithium burning} 
The lithium test is a primary method for confirming that an object is a brown dwarf \cite{1992ApJ...389L..83R}.~Observing lithium lines in stellar spectra allows astronomers to trace the core temperature history of young stars and brown dwarfs, distinguishing between different evolutionary epochs.~Therefore, deviations from the SM expectation of the lithium-7 abundance may provide a method for detecting dark dwarfs.~

Figure \ref{fig:Tcplot5} shows that for objects close to the stellar mass boundary the core temperatures of DM powered objects is cooler than the SM prediction.~The reaction that depletes lithium 
has a threshold temperature of $ \sim 2.5 \times 10^6 \rm \, K$.
~This suggests that dark dwarfs will have suppressed lithium burning rates.~At lower DM densities, we find a slight increase in core temperature, resulting in a slightly increased depletion rate.

To calculate the surviving light element fraction we consider the relevant reaction in this temperature range
\begin{align}
    ^7\textrm{Li} + \textrm{p} &\rightarrow ^4\textrm{He} + ^4\textrm{He}.~ 
  \label{eq:Liburn}
\end{align}
Objects with masses lighter than $0.35\msun$ are fully convective \cite{Chabrier:1997vx}, implying that the ratios of lithium to hydrogen are constant throughout the star.~The depletion rate is then given by \cite{Bildsten:1996ht,Ushomirsky:1997dw}
\begin{equation}
    \frac{d \ln f}{dt} = -\frac{ 4 \pi X}{M m_p} \int_0^R \rho^2 \langle \sigma v \rangle r^2dr
\label{eq:liEQN}
\end{equation}
where $X$ is the hydrogen mass fraction, $m_p$ is the proton mass, $f$ is the lithium to hydrogen ratio, and $ \sigma$ the relevant cross section, which we take from \cite{caughlan1988thermonuclear}.~We can solve Eq.~\eqref{eq:liEQN} using our analytic model but, because the nuclear burning rate \eqref{eq:Liburn} is highly sensitive to the core temperature, the model needs to be refined to include the corrections from Coulomb pressure in the stellar core.~We explain our method for this in Appendix~\ref{app:polytropic_approximation_failure}.

\begin{figure}[t!]
    \centering
    \includegraphics[width=0.9\linewidth]{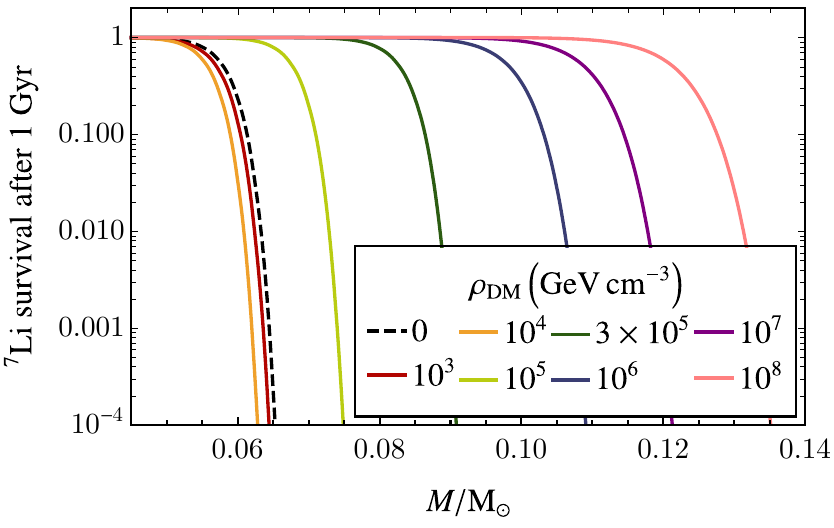}
    \caption{Lithium survival as a function of mass for different DM densities.}
    \label{fig:survivingLitDeu}
\end{figure}

In Fig.~\ref{fig:survivingLitDeu} we show the resulting survival fraction after 1 Gyr;~\vtwo{results for other benchmark ages can be computed using our accompanying code}.~In the case of the SM, we find that Li-7 is depleted in objects heavier than $0.062\msun$, consistent with other theoretical predictions \cite{1991MmSAI..62..171P,1993ApJ...413..364N,1996ApJ...459L..91C,2020A&A...637A..38P}.~In contrast, a significant fraction of lithium-7 survives in dark dwarfs of this mass and heavier in regions where $\rho_{\rm DM}\gtrsim 10^{5}$ GeV/cm$^3$.~Therefore, given an observed mass and estimated age, the spectroscopic detection of lithium can serve as a marker for DM heating.

\vtwo{Individual systems containing sub-stellar objects that are devoid of hydrogen burning lines are potential sites for detecting dark dwarfs, but an age estimate is also required to distinguish dark dwarfs from young brown dwarfs that have not yet burned their lithium.~In practice, a more promising approach would be to apply Bayesian hierarchical modeling e.g., the scheme of \cite{Benito:2024yki}, to use data from multiple systems to search for a statistical preference for the presence of dark dwarfs.} 

\section{Conclusion}
We have found that the structure, evolution, and fate of sub-stellar objects in the presence of dark matter annihilation is markedly different than the Standard Model predictions.~Objects lighter than the hydrogen burning limit begin their lives as brown dwarfs but ultimately evolve to become \textit{dark dwarfs} --- eternal objects powered by DM burning.~Their radii, temperatures, and luminosity are constant.~The H-burning limit itself is modified, depending on the incident DM flux.~Stars heavier than this behave similarly to red dwarfs but are predicted to be larger and brighter than SM red dwarfs of identical mass due to the additional contribution from DM heating.~Their core temperatures can be reduced, causing them to retain lithium in mass ranges where the SM predicts it is depleted.~The detection of lithium-7 in objects heavier than the lithium burning limit would provide evidence for the existence of DM heating.~The optimum location to search for these signatures is \vtwo{towards the galactic center, where low DM velocities and large densities maximize the incoming DM flux, with experiments such as JWST (which was shown to have sensitivity to brown dwarfs with temperatures as low as 650 Kelvin~\cite{Leane:2020wob} in such locations)}.

\section*{Software}

Mathematica version 12.0.0.~All of our results can be reproduced using our code, which is available at the following URL:~\href{https://zenodo.org/records/13141908}{https://zenodo.org/records/13141908}.

\section*{Acknowledgements}
DC is supported by the STFC under Grant No.~ST/T001011/1.~This material is based upon work supported by the National Science Foundation under Grant No.~2207880.

\appendix

\section{Appendix}

In this appendix we review the derivation of the analytic model that we adopted for our calculations above, explore the effects of the assumptions we have made, and derive the range of validity of our model.

\subsection{Polytrope Model for the Interior}

We begin by reviewing the polytropic description of sub-stellar objects, following reference \cite{2016AdAst2016E..13A}.~The EOS for the combination of electron degeneracy pressure $P_{\rm deg}$ and the gas pressure of the ions $P_{\rm ion}$ is an $n=3/2$ polytrope:\footnote{\vtwo{The general polytropic equation of state is given by:
$$
P = K \, \rho^{\frac{n+1}{n}},
$$
where the polytropic constant $K$ must be calculated from the microphysics. 
}
}
\begin{equation}
    P=K\rho^{\frac53}\textrm{ with }  K=C\mu_e^{-5/3}(1+\gamma(\psi)+\alpha\psi).
\end{equation}
In the above, $\mu_e=(X+Y/2)^{-1}$ is the number of electrons per baryon (with $X$ and $Y$ the mass fractions of hydrogen and helium respectively);~$\alpha=5\mu_e/2\mu$ with $\mu=((1+x_H)X+Y/4)$ the mean molecular weight of the ionized hydrogen and helium mixture ($x_H$ is the fraction of ionized hydrogen);
\begin{equation}
    C=\frac25 aA^{\frac52},~A=\frac{(3\pi^2\hbar^3N_A)^{\frac23}}{2m_e},~a=\frac23\frac{4\pi m_e^{\frac32}}{(2\pi \hbar)^3};
\end{equation}
and the \textit{degeneracy parameter}
\begin{equation}
\label{eq:psi_def}
    \psi=\frac{k_B T}{E_F}
\end{equation}
with $E_F$ the Fermi energy.~Numerically, $C=10^{13}$ cm$^4$g$^{-2/3}$s$^{-2}$.~The benchmark model in this work fixed $\mu_e=1.143$, and $\mu=1.23$ corresponding to a neutral mixture of 75\% hydrogen and 25\% helium;~this gives $\alpha=2.32$.
The function $\gamma(\psi)$ is given by
\begin{equation}
\label{eq:gammaeq}
    \gamma(\psi)=-\frac{5}{16}\psi\ln(1+e^{-\frac{1}{\psi}})+\frac{15}{8}\psi^2\left[\frac{\pi^2}{3}+\mathrm{Li}_2(-e^{-\frac{1}{\psi}})\right]
\end{equation}
with $\mathrm{Li}_2$ the polylogarithm of order-2.
Using the properties of $n=3/2$ polytropes, it is possible to find expressions for the radius, central temperature, central pressure, and central density as a function of the degeneracy parameter $\psi$ \cite{2016AdAst2016E..13A}:
%
\begin{align}
\label{eq:Rpoly}
    R&=R_0\mu_e^{-5/3}(1+\gamma(\psi)+\alpha\psi)M_0^{-\frac13}\,,\\
T_c&=T_0M_0^{\frac43}\mu_e^{\frac83}\frac{\psi}{(1+\gamma(\psi)+\alpha\psi)^2} \,,\\
    \rho_c&= \rho_0M_0^2\frac{\mu_e^{5}}{(1+\gamma(\psi)+\alpha\psi)^3} \,,\\
    P_c&=P_0M_0^{\frac{10}{3}}\frac{\mu_e^{\frac{20}{3}}}{(1+\gamma(\psi)+\alpha\psi)^4} \,,
    \label{eq:Pc}
\end{align}
where $M_0=M/\msun$,  $R_0= 
2.80858\times10^9$ cm, $\rho_0=  
1.28412\times10^5$ g/cm$^3$, $T_0  = 
7.68097\times10^8$ K, and $P_0= 
3.26763\times10^{12}\textrm{g/cm/s}^2$.~To make further progress, a model for the photosphere must be specified.

\subsection{Photosphere Models}

\begin{table}[h]
    \centering
    \begin{tabular}{c c c c}\hline
       Model & $b_1$ & $\nu$ & $2X_{H^+}$ \\\hline
       A  & 2.87 & 1.58 & 0.48 \\
       B  & 2.70 & 1.59 & 0.50 \\
       C  & 2.26 & 1.59 & 0.50 \\
       D  & 2.00 & 1.60  & 0.51 \\
       E  & 1.68 & 1.61 & 0.52 \\
       F  & 1.29 & 1.59 & 0.50 \\
       G  & 0.60 & 1.44 & 0.33 \\
       H  & 0.40 & 1.30 & 0.18
       \\\hline
    \end{tabular}
    \caption{Parameter points for the phase transition \cite{chabrier1992molecular}.~Note that points G and H are very close to the second critical point of hydrogen, where a liquid-liquid phase transition occurs.} 
    \label{tab:models}
\end{table}

The photosphere, which defines the effective temperature and the stellar luminosity, is the point at which the optical depth $\tau$ falls to $2/3$.~At this point, there is a phase transition from the mixture of degenerate electrons and ionic gas described above to a phase of molecular hydrogen.~Modeling the latter, one can derive expressions for the effective temperature and luminosity.~Several parameter points given in table~\ref{tab:models} span the possible range of surface temperatures at which the phase transition takes place:
\begin{equation}
    T_{\rm eff}= b_1 \times 10^6 \rho_e^{2/5} \psi^\nu \rm K,
\end{equation}
where $b_1$ and $\nu$ are parameters of the model \cite{chabrier1992molecular}.~Solving the condition $\tau=2/3$ yields an expression for the pressure at the photosphere, which can be used to find the density and effective temperature from the idea\vtwo{l} gas law (see \cite{2016AdAst2016E..13A} for the details -- here we generalise the solution presented in this work).~The result is:
\begin{align}
    T_{\rm eff}= 1.57466 \times 10^4
    \, b_1  \psi ^{\nu } \left(\frac{M_0^{5/3} \psi ^{-\nu }}{b_1 \kappa_R (1+\gamma+\alpha \psi)^2}\right)^{2/7}
    \rm K,
\end{align} 
where $\kappa_R$ is the Rosseland mean opacity.~Using the Stefan-Boltzmann law yields the surface luminosity, 
\begin{equation}
    \frac{L_{\rm surf}}{L_\odot}= \frac{0.0578353 b_1^3 M_0}{\kappa_R} \psi ^{3 \nu } \left(\frac{M_0^{5/3} \psi ^{-\nu }}{b_1 \kappa_R (1+\gamma+\alpha \psi  )^2}\right)^{1/7}.
\end{equation}
This equation is starting point for our work above.

\subsection{Sensitivity to the Molecular Hydrogen Phase Transition}

\begin{figure}[t!]
    \centering
    \includegraphics[width=0.85\linewidth]{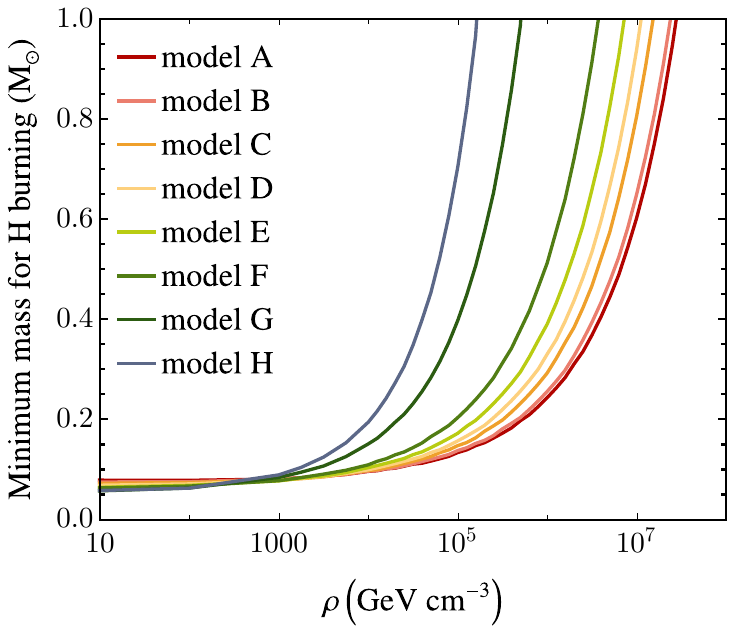}
    \caption{
  Minimum mass for hydrogen burning (using the criterion $L_{\rm DM} = L_{\rm HB}$) as a function of DM ambient density for different models of the photosphere.~Here we use  $ v_{\rm DM}=$50 km/s.}
    \label{fig:min_mass_hybrid}
\end{figure}

Our calculation relies upon a choice of parameter point for the phase transition, which introduces a source of theoretical uncertainty into our calculations.~We explore this in this section.~A convenient measure for quantifying the model-dependency is the minimum mass for hydrogen burning, which is DM density and velocity dependent.~Since DM annihilation reduces the core temperature and density below the threshold for PP-burning, the MMHB is increased whenever DM burning is important.~Demanding $L_{\rm HB}= L_{\rm DM}$ leads to the minimum mass shown in figure \ref{fig:min_mass_hybrid} for the different parameter points given in table~\ref{tab:models}.~The figure demonstrates that the model's predictions only significantly diverge for very large DM densities (with the exception of edge cases G and H), implying that the choice of parameter point is not an important  source of uncertainty for our conclusions.

\subsection{Coulomb Corrections to the Equation of State}
\label{app:polytropic_approximation_failure}

At low temperatures, the effects of Coulomb repulsion becomes important and the EOS deviates from an $n=3/2$ polytrope.~This has two consequences for our work.~First, it limits the applicability of the analytic model we have assumed to high mass objects and, second, the exponential sensitivity of the lithium-7 burning rate in Eq.~\eqref{eq:liEQN} mandates that we include these corrections to calculate the core temperature accurately.~We derive the mass range where our results are valid and expressions for the corrected core temperature in this section.

\begin{figure}
    \centering
    \includegraphics[width=0.85\linewidth]{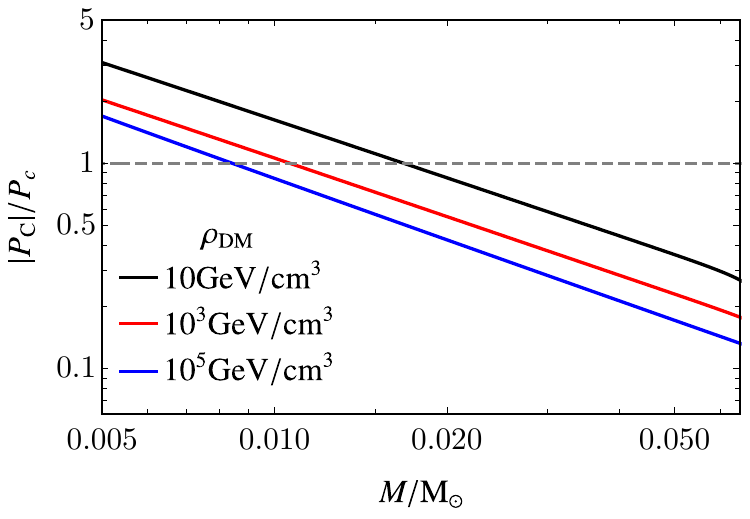}
    \caption{Ratio of the Coulomb to central pressure as a function of mass for varying DM densities.~The polytrope model presented in this work becomes invalid when $|P_{\rm C}|/P_c >1$.}
    \label{fig:validity}
\end{figure}

The effects of Coulomb interactions depend upon the plasma parameter \cite{1983psen.book.....C}
\begin{equation}
    \Gamma=\frac{e^2}{k_B T}\left(\frac{4\pi \rho }{3 m_p}\right)^{\frac13},
\end{equation}
where $e$ is the electron charge and $m_p$ is the proton mass.~Early in the star's evolution when lithium is being depleted the plasma parameter $\Gamma\ll 1$, corresponding to the  Deybe-H\"{u}ckel regime \cite{Ushomirsky:1997dw}.~In this limit, each ion can be thought of as being surrounded by a spherically symmetric but inhomogeneously charged cloud that screens its charge.~The Coulomb pressure of these screened charges is \cite{1992ApJ...389L..83R}
\begin{equation}
    P_{\rm C}=-\frac{e^2}{3}\left(\frac{\pi}{k_B T}\right)^{\frac12}\left(\frac{\rho\zeta}{m_p}\right)^{\frac32};~\textrm{with }\zeta=\sum_{j}\frac{X_j}{A_j}Z_j(1+Z_j),
\end{equation}
where $j$ runs over all ions with mass fraction $X_j$, atomic number $A_j$ and charge $Z_j$.~$\zeta=1.875$ for our benchmark model.~The ratio of the Coulomb pressure to the central pressure calculated using the polytrope model for various DM densities is shown in figure~\ref{fig:validity}.~In the SM, the polytropic approximation ceases to be valid ($|P_{\rm C}|>P_c$) when $M<\vtwo{0.018}\msun$.~This threshold is reduced when DM annihilation is present.~The results we presented in this work correspond to heavier objects.~

To incorporate the effects of the Coulomb pressure into our lithium calculation, we use that fact that $T_c\propto \mu_{\rm eff}^2$ \cite{Bildsten:1996ht,Ushomirsky:1997dw} with $\mu_{\rm eff}= k_B N_A \rho_c T_c/P_c$.~At fixed central density, the negative Coulomb pressure reduces the total pressure from the polytrope model's prediction, which raises the central temperature.~To incorporate this effect, we first calculate $T_c$ using the polytropic model and then scale this by a factor of $P_c^2/(P_c+P_{\rm C})^2$ to account for the correction to $\mu_{\rm eff}$.

At late times, when the star has cooled and reached its equilibrium state, either red or dark dwarf, the plasma parameter has evolved to $\Gamma\gg1$ and the Debye-H\"{u}ckel approximation does not apply.~Instead, the plasma is comprised of a sea of degenerate electrons surrounding a lattice of ions arranged to maximize their separation, minimizing their Coulomb repulsion \cite{1983psen.book.....C}.~In this limit, the Coulomb pressure can be found using the the Wigner-Seitz approximation where each ion with atomic number $Z$ sits at the center of a neutral sphere containing $Z$ electrons.~The Coulomb pressure is \cite{1983psen.book.....C}
\begin{equation}
\label{eq:Coulomb_Pressure}
\begin{split}
    P_{\rm C}&=K_{\rm C}\rho^{\frac43};\\ K_{\rm C}&=-\frac{3}{10}\left(\frac{4\pi \langle Z^2\rangle}{3}\right)^{\frac13}e^2\left(\frac{N_A }{\mu_e}\right)^{\frac43}\approx 5.72\times10^{12}\textrm{g/cm/s}^2.
\end{split}
\end{equation}
Comparing the central Coulomb pressure to the central pressure \eqref{eq:Pc}, the latter exceeds the former in objects where $M\lesssim 10^{-3}\,\msun \sim \rm M_{\rm jupiter}$, independent of the DM density.~Thus, the equilibrium configurations we have studied are valid at even lower masses than estimated above, the caveat being that the polytrope model provides an invalid description of their formation.

\clearpage
\bibliographystyle{JHEP}
\bibliography{refs}

\end{document}